\documentclass{article}
\usepackage{spconf,amsmath,graphicx,booktabs}
\usepackage{stfloats}
\usepackage{todonotes}

\def\L{{\cal L}}

\title{Exploring Attention Mechanism for Acoustic-based Classification of Speech Utterances into System-Directed and Non-System-Directed}
\name{Atta Norouzian$^1$, Bogdan Mazoure$^2$\sthanks{The author performed the work while an intern at Nuance}, Dermot Connolly$^1$ and Daniel Willett$^1$}  
\address{$^1$Nuance Communications\\ $^2$McGill University, Canada\\
\{atta.norouzian, dermot.connolly, daniel.willett\}@nuance.com, bogdan.mazoure@mail.mcgill.ca}
\begin{document}
\ninept
\maketitle
\begin{abstract}
Voice controlled virtual assistants (VAs) are now available in smartphones, cars, and standalone devices in homes. In most cases, the user needs to first ``wake-up'' the VA by saying a particular word/phrase every time he or she wants the VA to do something. Eliminating the need for saying the wake-up word for every interaction could improve the user experience. This would require the VA to have the capability to detect the speech that is being directed at it and respond accordingly. In other words, the challenge is to distinguish between system-directed and non-system-directed speech utterances. In this paper, we present a number of neural network architectures for tackling this classification problem based on using only acoustic features. These architectures are based on using convolutional, recurrent and feed-forward layers. In addition, we investigate the use of an attention mechanism applied to the output of the convolutional and the recurrent layers. It is shown that incorporating the proposed attention mechanism into the models always leads to significant improvement in classification accuracy. The best model achieved equal error rates of 16.25\% and 15.62\% on two distinct realistic datasets. 
\end{abstract}
\section{Introduction}
\label{sec:intro}
Thanks to recent advances in speech recognition and natural language understanding, VAs have become part of our daily lives. The VAs are typically activated by a wake-up word/phrase such as \textit{hi Mercedes}, \textit{hey BMW}, \textit{hey Siri}, \textit{Alexa} or \textit{ok Google}. Eliminating such wake-up words in favor of allowing direct requests for assistance from the VA could significantly improve the user experience. This requires the device to have the capability to detect speech directed at it and ignore human-to-human and background speech. The problem of classifying spoken utterances into system-directed and non-system-directed has previously been investigated within the context of virtual assistants~\cite{mallidi2018device, reich2011real, yamagata2009system} and dialogue systems \cite{wang2013understanding, dowding2006you}.  

Both, the spoken words and the way they are spoken provide cues for differentiating between system-directed and non-system-directed speech utterances. Typically, the lexical cues are extracted from a word sequence generated by an automatic speech recognition (ASR) system. The classification can then be performed by applying two language models \cite{shriberg2012learning, ravuri2014neural}, one for each class, to the hypothesized word sequence to compute a likelihood ratio and choose the class label based on that. Alternatively, the word sequence can be input to a neural network (NN) model to either directly estimate class probabilities \cite{ravuri2015recurrent} or to generate new features for another model \cite{mallidi2018device}. 
The non-lexical acoustic cues can be learned from features corresponding to prosodic structure \cite{reich2011real,shriberg2012learning} or the short-time frequency representation of the speech signal \cite{mallidi2018device, yamagata2009system}. The frequency-based features are typically extracted using a sliding window of 20-25 ms. One of the challenges involved in using frame-based features for utterance classification is to represent all utterances with a fixed-dimensional vector to train a model regardless of the length of the utterance. Averaging the features over time \cite{yamagata2009system} or passing the input sequence into a long short-time memory (LSTM) cell and using the last output of the LSTM cell~\cite{mallidi2018device} are examples of how others have dealt with this issue. In this paper, we propose a new technique based on attention to address some shortcomings of the other methods.

Similar to systems developed in \cite{mallidi2018device, reich2011real, yamagata2009system, shriberg2012learning, shriberg2013addressee} our plan is to combine information extracted from acoustic features with lexical information for this classification task, however, the focus of this paper is only on acoustic-based classification. The acoustic features explored for this purpose are frame-based log Mel-filterbank coefficients. We favor short-term frame-based acoustic features for this task since they facilitate early detection of user's intent by gradual application of the trained model on the incoming speech. The proposed classification models are based on deep neural networks (DNN) with combination of convolutional, recurrent and feed-forward layers. In addition, the use of an attention mechanism on top of the convolutional layer as well as the recurrent layer is investigated. 

This paper is organized as follows. First, an overview of the models developed for this classification problem is presented in \ref{sec:sys_overview}. A number of model architectures proposed for frame-based approach and the architecture of the utterance-based model are described in Section~\ref{sec:models}. The experimental study containing a description of the evaluation data, the model parameters, and the experimental results is provided in Section \ref{sec:exp}. Summary and conclusion are given in Section \ref{sec:conclusion}. 

\section{Overview}
\label{sec:sys_overview}
\begin{figure}[htb]
  \centering
  \centerline{\includegraphics[width=\linewidth]{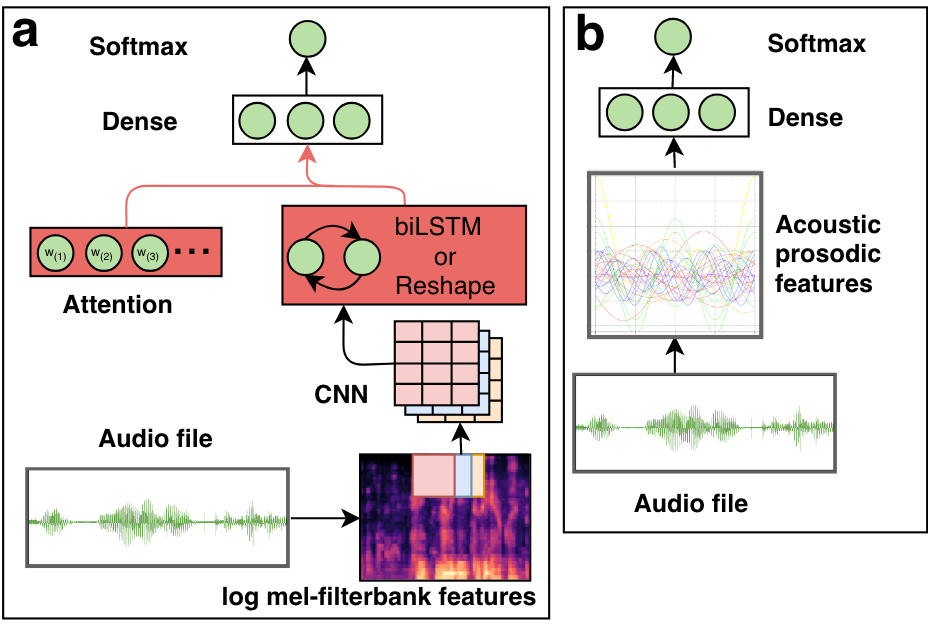}}
\caption{General architecture of frame-based models \textbf{(a)} and the utterance-based model \textbf{(b)} developed for the classification task at hand.}
\label{fig:model_architecture}
\end{figure}

Here, an overview of the two modeling approaches investigated for the system-directed versus non-system-directed classification problem is given. The two approaches are based on using frame-level and utterance-level input features. The models developed based on the frame-level input features are depicted in Figure ~\ref{fig:model_architecture} on the left and the utterance-based model in shown on the right. Several architectures are explored for the frame-based models as described in Section \ref{sec:models} for dealing with the variable length input sequences. The model developed for the fixed-length utterance-level features is comprised of only dense feed-forward layers. 

\section{Model Architectures}
\label{sec:models}
This section presents a number of model architectures for dealing with the issue of variable-length input feature representation faced in the frame-based approach. Moreover, it describes the input features and the architecture of the utterance-based model in detail.  

\subsection{Frame-based Approach}
\label{sub_sec:frame_based}
In this approach the feature vectors input to the models consist of 45 log Mel-filterbank coefficients extracted from 25 ms of acoustic signal with a frame shift of 10 ms. A speech utterance is hence represented by a sequence of feature vectors, $\{\boldsymbol{m}_1^{45},\ldots,\boldsymbol{m}_T^{45}\}$, where ${T}$ is the total number of frames in the utterance. All frame-based models developed here use a two-dimensional convolutional layer as input layer which outputs a set of ${d}$ feature maps denoted by $\{\boldsymbol{E}_1^{j\times l},\ldots,\boldsymbol{E}_d^{j\times l}\}$. The width of the feature maps, ${j}$, is proportional to the acoustic feature vector size (i.e., 45) and their length ${l}$ is proportional to ${T}$. In a realistic scenario, recorded utterances have different lengths which means ${T}$ and consequently ${l}$ vary from one utterance to another. This causes an issue when converting the feature-maps into a vector to pass to feed-forward layers since the input to a feed-forward layer has to have a fixed-dimension for all samples. In the following, three approaches for creating a fixed-length vector from variable-length feature-maps are presented. After creating a fixed-length vector it is input to dense feed-forward layers followed by a softmax layer as shown in Figure ~\ref{fig:model_architecture}.

\textbf{Global averaging across time:} A simple way of generating a fixed-length representation is to take the average of each feature-map over its length ${l}$. This will transform every 2-D feature map $\boldsymbol{E}_i^{j\times l}$ to a vector ${\boldsymbol{e}_i^j}$. The resulting vectors, $\{\boldsymbol{e}_1^{j},\ldots,\boldsymbol{e}_d^{j}\}$, are then concatenated and fed to a feed-forward layer as was done in \cite{shon2018convolutional}.

\textbf{Using a recurrent layer:} One could obtain a fixed-length vector from variable-length feature maps by using a recurrent neural layer. This is done by first concatenating columns of all feature maps to generate ${l}$ super vectors of dimension ${d \times j}$. Next, the super vectors are fed to a recurrent layer one by one and the last (i.e., ${l}$th) output of the recurrent layer is used for the succeeding layer. In addition, one could use a bi-directional recurrent layer and use the last output vector of forward and backward directions to obtain a richer fixed-length representation. In the model explored here, a bi-directional LSTM layer is used for this purpose and the two resulting vectors from both directions are concatenated and used in the feed-forward layer.

\begin{figure}
    \centering
    \includegraphics[width=0.9\linewidth]{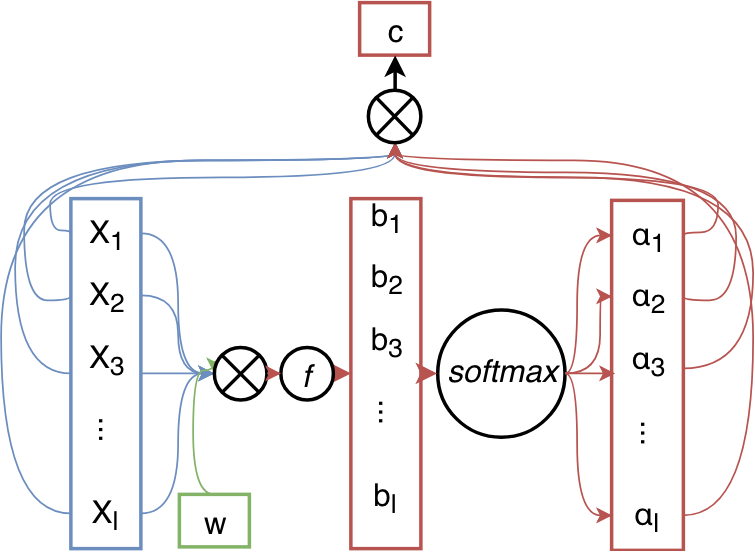}
    \caption{Attention mechanism applied to the input sequence ${\boldsymbol{X}}$ to generate the context vector $\boldsymbol{c}$.}
    \label{fig:my_label}
\end{figure}

\textbf{Using attention mechanism:} Simple averaging of feature maps or passing them through a recurrent layer and using only its last output could result in losing important information. An attention mechanism could retain most of the relevant information while resolving the variable-length issue. The attention mechanism explored here is somewhat different from the traditional encode-decoder based attention introduced in \cite{bahdanau2014neural}. It is in essence a weighted average of sequence of vectors where the weights are learned through back-propagation. This mechanism was first explored for emotion recognition in \cite{neumann2017attentive} and is similar to the idea of self-attention in \cite{vaswani2017attention}. Denoting a sequence of ${l}$ vectors of dimension ${s}$ by the matrix $\boldsymbol{X}^{s \times l}$, attention is computed as
\begin{equation}
    \begin{split}
        \boldsymbol{b}^{l\times 1} &= f(\boldsymbol{w}^{1\times s} \boldsymbol{X}^{s\times l}), \\
        \alpha^{i} &= \frac{\exp(b_i)}{\sum_{j=1}^l\exp(b_j)}, \ \ i = 1, \ldots, l, \\
    \end{split}
    \label{eq:attention}
\end{equation}
where $\boldsymbol{w}$ is the weight vector learned through back-propagation, ${f}$ is a non-linear function (here $tanh$), $b$ is the attention vector, and $\alpha$ is the normalized attention vector. Applying attention to the input sequence results in a vector known as context vector given by
\begin{equation}
\boldsymbol{c}^{s\times 1} = \boldsymbol{X}^{s\times l}\boldsymbol{\alpha}^{l\times 1}.
\label{eq:context}
\end{equation}
As can be seen in Equation \ref{eq:context}, the context vector dimension is independent of the length of the input sequence ${l}$. Furthermore, the attention vector ${\boldsymbol{\alpha}}$ helps to put more emphasis on the parts of the input sequence ${\boldsymbol{X}}$ that carry the most relevant information for distinguishing the two classes. The process of computing the attention and applying it to the input sequence is shown in Figure~\ref{fig:my_label}. The input sequence in this case could be the flattened feature maps or the output of the recurrent layer.

\subsection{Utterance-based Approach}
\label{sub_sec:utterance_based}
As an alternative to the frame-based approach one could represent every utterance with a fixed-length feature representation prior to any modeling. This can be done by computing some functions over the frame-based features. The feature set used here was developed for INTERSPEECH ComParE emotion recognition sub-challenge \cite{schuller2013interspeech}. It contains 6373 acoustic-features described in \cite{weninger2013acoustics}. We used the openSmile toolkit \cite{eyben2013recent} for extracting these features from the speech utterances in our corpus. Although, this feature set was originally developed for an emotion recognition task, it contains a variety of acoustic-prosodic features (F0, energy, zero crossing, mfccs) many of which are relevant for this classification problem as well. A three-layer dense feed-forward model is trained and evaluated for these features.
\begin{figure*}[htb]
  \centering
  \centerline{\includegraphics[width=0.9\linewidth]{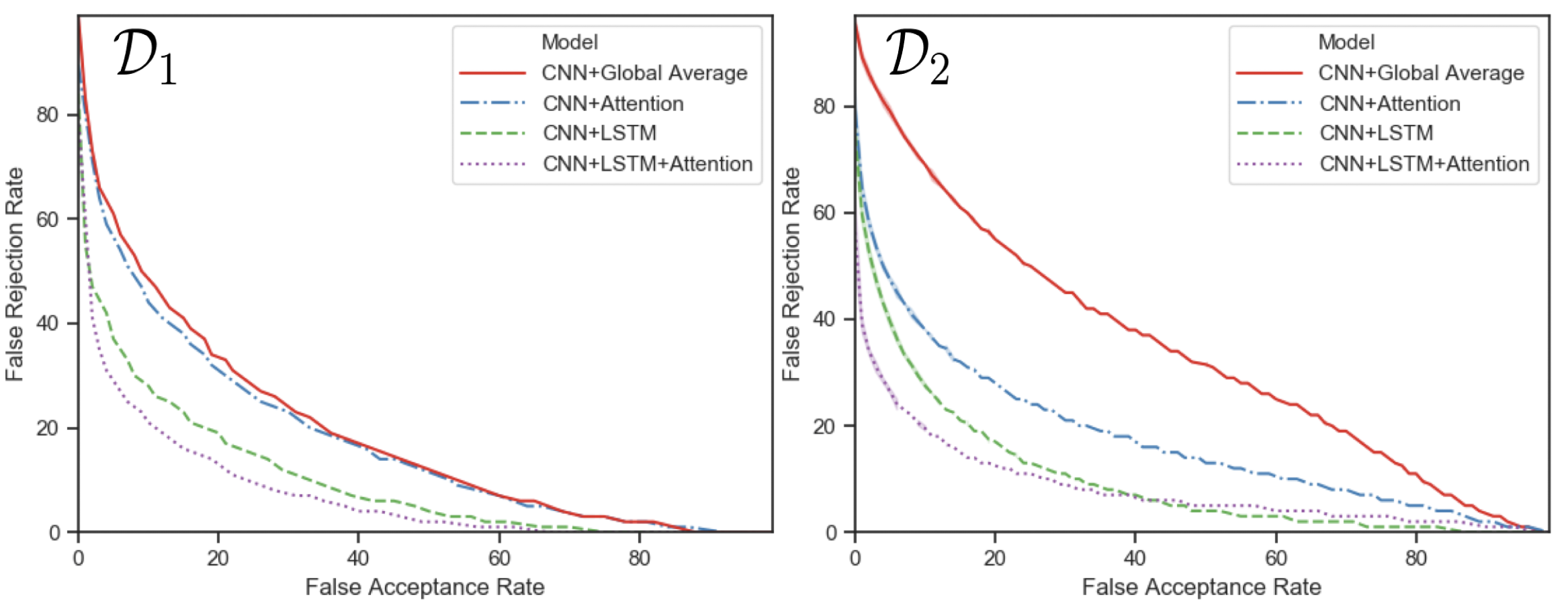}}
\caption{Detection error trade-off curves of the four frame-based models on dataset $\mathcal{D}_1$ (left) and dataset $\mathcal{D}_2$ (right).}
\label{fig:eer}
\end{figure*}

\section{Experimental study}
\label{sec:exp}
This section provides a description of the two data sets used for training and evaluation of the classifier. Afterwards, the model parameter are given and finally experimental results are presented and analyzed. 

\subsection{Datasets}
\label{sub_sec:dataset}
Two datasets are used for evaluation of the proposed techniques. The first dataset denoted by $\mathcal{D}_1$ contains recordings from a device with virtual assistant. The recordings contain system-directed utterances including questions and commands as well as non-system directed utterances mostly consisting of people dictating phrases. These are actual recordings from multiple users in different environments. The second dataset denoted by $\mathcal{D}_2$ also contains virtual assistant recordings but in addition it includes, background speech, open microphone recordings and some non-speech noise in the non-system-directed subset. The models are only trained on the training subset of $\mathcal{D}_1$ and the dataset $\mathcal{D}_2$ was only used for testing. Table~\ref{tab:dataset_breakdown} shows the breakdown of both datasets by class and training/validation/test. Same number of training utterances from both classes where chosen for training the models to prevent them from being biased towards one class. 

\begin{table}[h]
    \centering
    \begin{tabular}{lcccc}  
\toprule
{} & \multicolumn{3}{c}{$\mathcal{D}_1$} & $\mathcal{D}_2$ \\
\cmidrule(r){2-4}
& Training    & Validation & Test & Test \\
\midrule
System & 35k & 12k & 14k & 7k \\
Non-system & 35k & 5k & 4k & 127k \\
\bottomrule
\end{tabular}
    \caption{Number of utterances per training, validation and test splits across both classes and datasets $\mathcal{D}_1$ and $\mathcal{D}_2$ rounded to the nearest thousand.}
    \label{tab:dataset_breakdown}
\end{table}

\subsection{Model Parameter}
\label{sub_sec:model_param}
Prior to training the models, the frame-based and utterance-based feature vectors are normalized to have zero mean and unit standard deviation along each dimension to facilitate model convergence. Moreover, Adam optimizer and early stopping are used for training the models. For the frame-based models, the convolutional layer has a depth of 50 with a kernel height of 20 and width of 9 for the models without the LSTM layer and width of 5 for the models with LSTM layer. A stride of 5 is used along the time access and 3 along the log Mel-filterbank coefficients. The LSTM layer is bi-directional with 128 units in each direction. The three feed-forward layers in frame-based models each have 128 units. The best classification accuracy was obtained from the utterance-based model when three feed-forward layers of 128 units were used. All the models were trained using tensorflow toolkit \cite{tensorflow2015-whitepaper}.

\subsection{Results}
\label{sub_sec:results}
In this section the classification models described in Section~\ref{sec:sys_overview} are evaluated on $\mathcal{D}_1$ and $\mathcal{D}_2$ datasets defined in Section~\ref{sub_sec:dataset}. Figure~\ref{fig:eer} shows the performance of the models in terms of detection error tradeoff (DET) curves. A number of observations can be made from these plots. First, using an LSTM significantly improves the model performance compared to global averaging. Moreover, adding attention to the mix yields an additional boost to the performance with and without the LSTM. Furthermore, having larger improvement with attention on $\mathcal{D}_2$ dataset which was not seen in training suggests that the proposed attention mechanism improves model generalization as well. To have a single point of reference to compare the models, the equal error rate metric which corresponds to equal Type I and Type II errors is measured and shown in Table~\ref{tab:eer}. 
\begin{table}[htb]
    \centering
    \begin{tabular}{lcc}  
\toprule
& $\mathcal{D}_1$ & $\mathcal{D}_2$ \\
\midrule
CNN+Global Average & 26.99 & 39.25\\
CNN+Attention & 26.21 & 24.90\\
CNN+LSTM & 19.46 & 18.79\\
CNN+LSTM+Attention & \textbf{16.25} & \textbf{15.62}\\
\bottomrule
\end{tabular}
    \caption{Equal error rates of the four frame-based models measured on $\mathcal{D}_1$ and $\mathcal{D}_2$ test sets.}
    \label{tab:eer}
\end{table}

\begin{figure}[htb]
  \centering
  \centerline{\includegraphics[width=\linewidth]{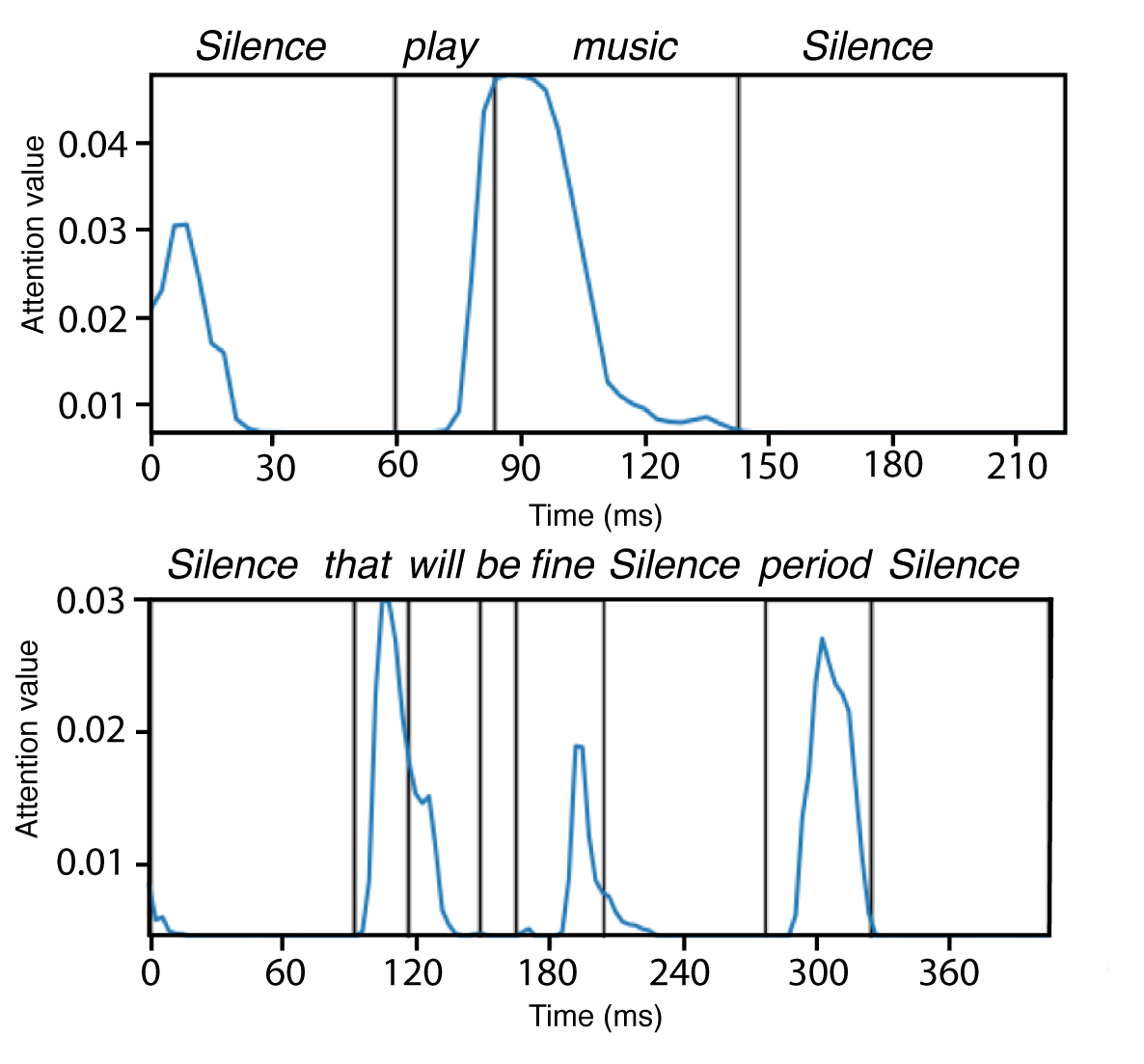}}
\caption{Attention vector spread across time for a system-directed utterance (top) and a non-system-directed utterance (bottom) from $\mathcal{D}_1$ testset.}
\label{fig:attention}
\end{figure}
The main question here is what the model is actually learning. This is not easy to answer especially when it comes to neural network models. However, the attention mechanism could help shedding some light on this matter. Aligning the attention vector ${\boldsymbol{\alpha}}$ with the original speech utterance, one could find out where the model is putting the most emphasis. This is done in Figure~\ref{fig:attention} for two utterances from the two classes. The word sequences associated with the utterances are also shown in the figure to identify possible correlations between the spoken words and where the model is mostly focusing on. The vertical lines correspond to start and end time of the words. The plot shows that for the system-directed utterance the attention is on both ``\textit{play}'' and ``\textit{music}'' while for the non-system-directed example the attention is mostly on the words ``\textit{that}'', ``\textit{fine}'', and ``\textit{period}''. It is interesting to note that in the training dataset the word ``\textit{period}'' is spoken only when the users are dictating a phrase. In other words, this word is a strong indication that the speech utterance is of dictation style which belongs to non-system-directed class. This led us to think that maybe the model is just learning keywords and is not learning any para-linguistic information. To answer this question we looked at a number of system-directed utterances such as ``\textit{you didn't catch that}'' and ``\textit{one more run after that}'' that were not part of the training data and did not contain any word highly correlated with system-directed class. The model classified both of these utterances correctly with high confidence. This suggests that the model is not just learning keywords or para-linguistic information but rather a combination of both. Adding more training data from different domains would make the model less sensitive to words and more sensitive to para-linguistic information.

\begin{figure}[htb]
  \centering
  \centerline{\includegraphics[width=\linewidth]{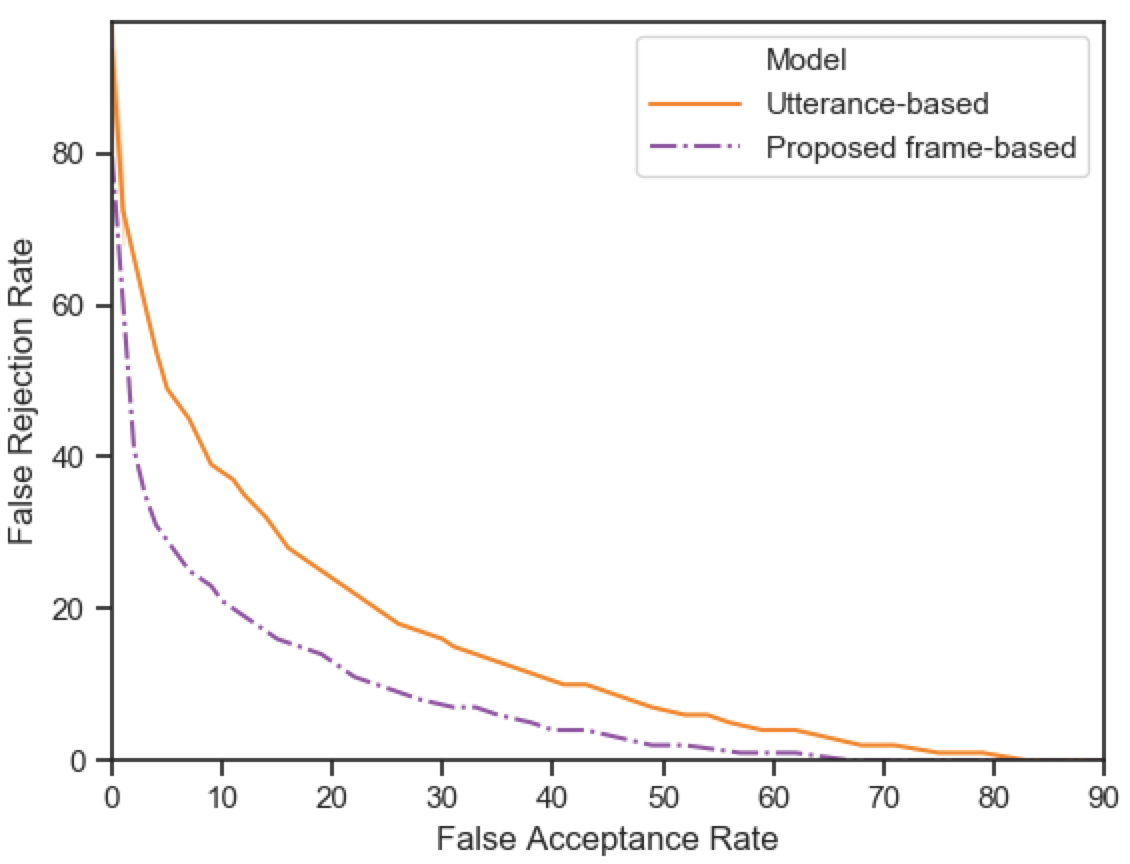}}
\caption{Detection error trade-off curves for the utterance-based approach and the proposed frame-based approach measured on $\mathcal{D}_1$ testset.}
\label{fig:compare}
\end{figure}
In Figure~\ref{fig:compare}, the proposed frame-based approach is compared to the utterance-based approach described in Section~\ref{sub_sec:utterance_based} on the $\mathcal{D}_1$ dataset. It should be noted that the utterance-based acoustic-prosodic feature set was designed for emotion recognition and contains several features that may not be relevant for this task. Nevertheless, the gap between the two curves indicates that even without using hand-crafted features and only relying on frame-based log Mel-filterbank features very good classification performance can be achieved with the proposed attention-based modeling technique.   

\section{Conclusion}
\label{sec:conclusion}
In this paper, the problem of classifying speech utterances into system-directed and non-system-directed was addressed. A number of neural network architectures  based on using convolutional and recurrent layers were investigated. It was shown that having an attention mechanism improves the classification performance whether applied directly to the output of the convolutional layer or to the output of the recurrent layer. The best performing model was built by stacking a convolutional layer, a recurrent layer and three feed-forward layers with attention applied to the output of the recurrent layer. This model achieved an EER rate of ${16.25\%}$ on one test set and ${15.62\%}$ on the second test set. As continuation of this work we are looking into combining direct audio classification with ASR-output based text classification for improved accuracy.

\section{Acknowledgements}
The authors would like to thank two colleagues from Nuance Communications, Raymond Brueckner for helping out with the openSmile toolkit and Yasser Hifny for very insightful discussions.

\bibliographystyle{IEEEbib}


\end{document}